\documentclass[sigconf]{acmart}

\usepackage{enumitem}
\usepackage{verbatim}
\usepackage{bm}
\usepackage{multirow}
\usepackage{multicol}
\usepackage{array}
\usepackage{booktabs}
\usepackage{nomencl}
\usepackage{appendix}
\usepackage{amsfonts}
\usepackage[marginal]{footmisc}

\AtBeginDocument{%
  \providecommand\BibTeX{{%
    \normalfont B\kern-0.5em{\scshape i\kern-0.25em b}\kern-0.8em\TeX}}}

\acmYear{2024}
\setcopyright{acmlicensed}
\acmConference[WWW '24 Companion] {Companion Proceedings of the ACM Web Conference 2024}{May 13--17, 2024}{Singapore, Singapore.}
\acmBooktitle{Companion Proceedings of the ACM Web Conference 2024 (WWW '24 Companion), May 13--17, 2024, Singapore, Singapore}
\acmISBN{979-8-4007-0172-6/24/05}
\acmDOI{10.1145/3589335.3648345}
\begin{document}

%%
%% The "title" command has an optional parameter,
%% allowing the author to define a "short title" to be used in page headers.
%\title{An End-to-End Interpretable Numerical Feature Learning Framework for Recommender System}

\title{Enhancing Interpretability and Effectiveness in Recommendation with Numerical Features via Learning to Contrast the Counterfactual samples}

\author{Xiaoxiao Xu}
\affiliation{%
  \institution{Kuaishou Technology}
  \country{Beijing, China}
}
\email{xuxiaoxiao05@kuaishou.com}

\author{Hao Wu}
\affiliation{%
  \institution{Kuaishou Technology}
  \country{Beijing, China}
}
\email{wuhao10@kuaishou.com}

\author{Wenhui Yu}
\affiliation{%
  \institution{Kuaishou Technology}
  \country{Beijing, China}
}
\email{yuwenhui07@kuaishou.com}

\author{Lantao Hu}
\affiliation{%
  \institution{Kuaishou Technology}
  \country{Beijing, China}
}
\email{hulantao@kuaishou.com}

\author{Peng Jiang}
\affiliation{%
  \institution{Kuaishou Technology}
  \country{Beijing, China}
}
\email{jiangpeng@kuaishou.com}

\author{Kun Gai}
\affiliation{%
  \institution{Unaffiliated}
  \country{Beijing, China}
}
\email{gai.kun@qq.com}

%%
%% The "author" command and its associated commands are used to define
%% the authors and their affiliations.
%% Of note is the shared affiliation of the first two authors, and the
%% "authornote" and "authornotemark" commands
%% used to denote shared contribution to the research.

%%
%% By default, the full list of authors will be used in the page
%% headers. Often, this list is too long, and will overlap
%% other information printed in the page headers. This command allows
%% the author to define a more concise list
%% of authors' names for this purpose.
\renewcommand{\shortauthors}{Xiaoxiao Xu et al.}

%%
%% The abstract is a short summary of the work to be presented in the
%% article.
\begin{abstract}
We propose a general model-agnostic \textbf{C}ontrastive learning framework with \textbf{C}ounterfactual \textbf{S}amples \textbf{S}ynthesizing (CCSS) for modeling the monotonicity between the neural network output and numerical features which is critical for interpretability and effectiveness of recommender systems. CCSS models the monotonicity via a two-stage process: synthesizing counterfactual samples and contrasting the counterfactual samples. The two techniques are naturally integrated into a model-agnostic framework, forming an end-to-end training process.
Abundant empirical tests are conducted on a publicly available dataset and a real industrial dataset, and the results well demonstrate the effectiveness of our proposed CCSS. Besides, CCSS has been deployed in our real large-scale industrial recommender, successfully serving over hundreds of millions users.
\end{abstract}

%%
%% The code below is generated by the tool at http://dl.acm.org/ccs.cfm.
%% Please copy and paste the code instead of the example below.
%%



%% A "teaser" image appears between the author and affiliation
%% information and the body of the document, and typically spans the
%% page.

%%
%% This command processes the author and affiliation and title
%% information and builds the first part of the formatted document.
\maketitle

\footnote{First Author and Second Author contribute equally to this work.}

\section{Introduction}
\begin{figure}[t]
\centering
\includegraphics[width=1.1\linewidth]{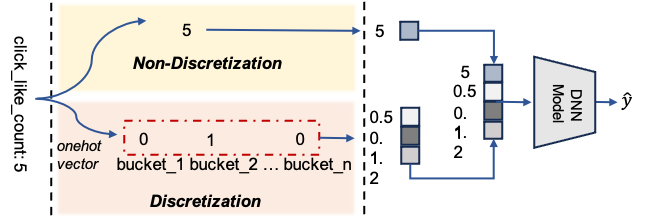}
\vspace{-1.0em}
\caption{An illustration of the numerical features usage in recommender's deep model: utilized both as dense values by Non-discretization approach and as embedding by Discretization approach.}
\label{illu_num_fea}
\vspace{-1.5em}
\end{figure}

Recommender systems equipped with deep models have been widely deployed industrially to alleviate the problem of information overload.
Deep models in industrial recommender systems, eg., CTR prediction model and likes prediction model, adopt numerical features as well as categorical features as inputs.
Numerical features learning has become an active research task in industrial recommender systems \cite{guo2021embedding,10.1145/3511808.3557587}. 
%such as Number of Clicks, averaged empirical CTR, averaged predicted CTR, etc,.
Illustrating the activity levels of users, items and the interactions between specific users and items, numerical features are crucial to the performance of deep models.
% TODO: 相比类别特征，数值类特征的特点，连续性和可比的大小关系
% numerical features有两种使用方式，一种是借助于离散化手段转化为embedding使用，另一种直接作为dense feature使用
In practice, learning approach for numerical features can be categorized into two groups: (1) Non-discretization: using the original values or its transformations directly as dense inputs \cite{non-dis1,non-dis2,non-dis3}; (2) Discretization: transforming continuous numerical features into categorical features through discretization strategy and assigning embedding like categorization strategy \cite{guo2021embedding}.
In general, the above two learning approaches for numerical features are both adopted in a single deep neural network, as illustrated in Figure \ref{illu_num_fea}.

Most of the existing research for numerical features focuses on discretization and representation techniques to retain continuity for similar features \cite{guo2021embedding,10.1145/3511808.3557587}, but the prior semantic relation, i.e., monotonicity between the neural network output and the numerical inputs has been seldom addressed publicly. 
The monotonicity between the model output and numerical inputs is critical to the interpretability of the neural network.
In an industrial Recommender system, with the same context, the higher the activity level of the interaction between the user and the item, the higher predicted score of the neural network. Especially in User-Generated Content based platforms, such as Tiktok, Kuaishou, and etc., the consumer expect to see UGC with high quality illustrated by numerical features. Besides, in UGC-based platforms, contents with high quality are rewarded by more chance to display, which is critical for motivating the producer.
% 介绍单调性，并强调单调性的重要，点出数值类特征与模型输出之间的单调性，也是一种可解释性，并指出这种可解释性的重要（对ugc平台很重要，历史表现好的ugc，模型打分高，预期获得流量的概率大）
% 指出单调性也是一种先验知识，显式建模出来，会提高模型准确度
Besides, the monotonicity between the neural network outputs and numerical inputs is one kind of prior knowledge of the Recommender system, it is helpful for improving the accuracy of the model.

Traditional methods for modeling the monotonicity between neural network output and the numerical inputs are of two types: (1) only adopt numerical inputs as dense features and fuse them with DNN output linearly \cite{mono-dense}; (2) only adopt numerical inputs as dense features with a delicately designed network to retain monotonicity \cite{mono-emb}. % 这里可以加两张小图示意
Above methods both require that numerical features should only be utilized as dense values through Non-discretization approach.
In a real industrial Recommender system, besides adopted as dense features by Non-discretization approach, numerical features are also learned as embedding by Discretization approach, as shown in figure 1, in which the traditional methods are no longer applicable.
% 介绍已有的numerical feas可解释性的相关工作（Airbnb），并指出它的不适用之处
% 提出我们的框架，并点出我们框架的优势

To explicitly model the monotonicity between the neural network outputs and numerical inputs and make better use of this prior knowledge, we propose an adoptable model-agnostic \textbf{C}ontrastive learning framework with \textbf{C}ounterfactual \textbf{S}amples \textbf{S}ynthesizing (CCSS). CCSS consists one Counterfactual Samples Synthesizer and one Contrastive Objective Function Module which can cooperatively serve as a plug-and-play component to any neural network. The Counterfactual Synthesizer generates a counterfactual sample and a factual sample by disturbing the value of numerical features and keeping other features unchanged of the original sample.
% 尽量使得合成样本与原始样本相似，这样可以有效利用contrastive loss的效用
To synthesize generated samples as realistic as possible, which can take advantage of the contrastive loss to the greatest extent, we disturb only one numerical feature for each original sample.
Moreover, to generate samples as informative as possible, we introduce feature importance as the specific metric to quantize the probability to be disturbed for each numerical feature.
The contrastive objective function learns to rank the model outputs of the generated sample and the original sample, which enable the model to learn the monotonicity between model outputs and numerical features in an end-to-end manner. We summarize the main contributions as follows:
\begin{itemize}[leftmargin=10pt]
\setlength{\itemsep}{-1pt}
    \item As far as we know, we are the first to clarify the importance to neural network's interpretability and effectiveness of the monotonicity between the neural network output and the numerical features.
    \item We propose a general model-agnostic Contrastive learning framework with Counterfactual Samples Synthesizing(CCSS). The counterfactual and factual samples are generated locally based on the original samples, which enable CCSS to be adoptable in online learning scheme. The contrastive loss among the counterfactual sample, the factual sample and the original sample makes it possible to model the monotonicity end-to-end.
    \item We verify the effectiveness and adaptability of our proposed CCSS through extensive experiments conducted on a public benchmark dataset and a large scale industrial dataset, and successfully apply it in a real-world large scale recommender system bringing a considerable performance improvement.
\end{itemize}

% CCS的原理
% 反事实样本生成（生成技术？按照特征重要性来生成反事实样本）
%    1. 按照特征重要性归一化，作为采样概率
%    2. delta 如何确定？
%       1) 超参数
%       2) 统计，按照什么来做统计呢？
% 对比loss，端到端建模单调性
% 

% 总结我们这篇论文的contribution

% essential for the representation of the numerical features to retain continuity for similar features as well as discriminability for distinct features.

\section{Related Work}
In this section, we will introduce the related work from three aspects: Interpretability in Recommendation, Numerical Features Modeling in Recommendation and Counterfactual Sample Synthesizing.

\subsection{Interpretability in Recommendation}
% 可解释性在推荐系统中意味着什么，为什么重要，有哪些手段提升推荐系统的可解释性。点明值预估模型中可解释性讨论很少
Recommender systems play a crucial role in a wide range of web applications and services, in terms of distributing online contents to targeted users who are likely to be interested in them. The majority of the efforts in recommender system have been put into developing more effective model structures to achieve better performance, while the interpretability, i.e., explainability of recommender system has been seldom addressed. The interpretability of recommender system gives reasons to clarify why such results are derived \cite{ribeiro2016should}, which benefits recommender system in to ways: 1) interpretability helps system maintainers to diagnose and refine the recommendation pipeline. 2) interpretability promote persuasiveness and customer satisfaction by increasing transparency \cite{liu2020explainable}. Research on interpretability in recommendation can be grouped into two directions \cite{zhang2020explainable}: giving recommendation reasons by post-hoc methods \cite{10.1145/3219819.3220072} and building explainable models or add explainable components to give reasonable recommendation \cite{liu2020explainable}.
Recently, the efforts in recommender's interpretability have been concentrated on introducing more structural information as model inputs \cite{SHIMIZU2022107970}. Our work falls into the second direction. In this paper, we propose to improve the interpretability of models in recommendation from the perspective of building the monotonicity between the outputs of the model and the numerical feature, in which the research is limited.

\subsection{Numerical Features Modeling in Recommendation}
% 推荐系统中，比如ctr模型等，数值类特征重多，并且很重要。目前推荐系统中数值类特征的建模方法有xxx，但是他们集中在为数值类特征学习更好的embedding表征，忽视了模型输出与树枝类特征之间的单调性先验。单调性先验意味着什么，为什么重要。
In recommendation, e.g., CTR prediction model, learning of numerical features is one of the hot research topics. Most of the work is aimed to learning better embedding for numerical features in Discretization approach by means of addressing SBD (Similar value But Dis-similar embedding) and DBS (Dis-similar value But Same embedding) problems \cite{guo2021embedding,guo2017deepfm,10.1145/3511808.3557587}. In this paper, we propose to learn the monotonicity between model outputs and numerical features, which is another type of prior information of numerical features.
It is helpful for deriving more reasonable recommendation to model the prior monotonicity between neural network outputs and numerical features. To explain this, we take a frequently-used numerical feature $click\_like\_count$ for instance. With other features being the same, the video with higher $click\_like\_count$ should be ranked higher than the other in the same context. Our work model the monotonicity end-to-end with an model-agnostic method, which is the first attempt in numerical feature learning research domain.

\subsection{Counterfactual Sample Synthesizing}
% 在CV、NLP领域中应用较多，有两种主要用途：数据增强，提升模型鲁棒性
% 在推荐系统中应用较少，主要难点在于样本生成，在本文中，我们在推荐系统中应用，并且使用active sampling的方式生成样本（如何与uncertainty结合？https://arxiv.org/pdf/2305.13535.pdf）
Counterfactual samples have been recently used for data augmentation in Visual Question Answering (VQA) and Natural Language Processing (NLP). 
% 给定一个original sample，改变其中一些特征，对应地改变它的label，得到一条counterfactual sample
In VQA, counterfactual samples are synthesized by masking the critical objects in images or words in questions \cite{Chen_2020_CVPR}, which is much adoptable than traditional adversary-based data augmentation methods.
In NLP, counterfactual samples are synthesized by removing phrases, which al% https://aclanthology.org/2021.naacl-main.18/  https://arxiv.org/pdf/2305.13535.pdf
The generated counterfactual samples have also been used for explanations of deep learning, i.e. interpretability \cite{verma2020counterfactual}.
In recommendation, we synthesize counterfactual samples by directionally disturbing numerical feature values of the original samples. For each sample, the numerical feature to be disturbed is sampled with its feature importance as sampling probability. In addition, to model the monotonicity between the model output and numerical features, we introduce auxiliary contrastive objective ranking the model outputs of the counterfactual sample and the factual sample.

\section{Proposed Method}
In Section 3.1, we firstly review the background of deep models in industrial recommender systems, using CTR prediction as an example for illustration. Then we describe how to synthesize counterfactual samples in our proposed framework in Section 3.2. We further dig into the details of the implementations during training in Section 3.3.
The notations are summarized in Table \ref{tab:notation}.

\begin{table}[]
    \centering
    \renewcommand\arraystretch{1.5}
    \begin{tabular}{l|l}
    \toprule
    \specialrule{0em}{1pt}{1pt}
    %\specialrule{0em}{1pt}{1pt}
        %$\bm{x}$ & feature vector of input instance\\
        %$y$ & observed label for input instance\\
        %$\hat{y}$ & estimated CTR for input instance\\
        %$\bm{z}$ & the input embedding of MLP module\\
        $\mathcal{O}$ & Original sample\\
        $\mathcal{F}$ & Factual sample\\
        $\mathcal{C}$ & Counterfactual sample\\
        $\bm{E}$ & embedding parameters\\
        $x_{i}^{c}$ & feature value of the i-th categorical feature field\\
        $x_{j}^{n}$ & feature value of the j-th numerical feature field\\
    \bottomrule
    \end{tabular}
    \caption{Important notations.}
    \label{tab:notation}
    \vspace{-2.5em}
\end{table}

\subsection{Preliminaries}
% 推荐系统中的值预估模型，二分类任务，是否以ctr任务为例，模型结构，特征使用，loss设计
%\subsubsection{CTR Prediction Problem Formulation}
Given an user, a candidate item and the contexts in an impression scenario, CTR prediction, is to infer the probability of a click event. The CTR prediction model is mostly formulated as a supervised logistic regression task and trained with an i.i.d. dataset $\mathcal{D}$ collected from historic impressions. Each instance $\bm{O}=(\bm{x}, y) \in \mathcal{D}$ contains the features $\bm{x}$ implying the information of ${\{user, item, contexts\}}$, and the label ${y \in \{0, 1\}}$ observed from user implicit feedback. $y=1$ indicates an instance with positive label, and an instance with negative label is indicated by $y=0$. The instance feature $\bm{x}$ is a multi-field data record including M numerical fields and N categorical fields:
\begin{equation}
\bm{x}=concatenate([x_{1}^{c},x_{2}^{c},...,x^{c}_M], [x^{n}_{1},x^{n}_{2},...,x^{n}_{N}])
\end{equation}

For the i-th categorical feature field, the feature embedding can be obtained by embedding look-up operation:
\begin{equation}
    \bm{e}^{c}_i=\bm{E}^{c}_i\cdot onehot(x^{c}_i)
\end{equation}
where $\bm{E}^{c}_i \in \mathbb{R}^{v_i \times d}$ is the embedding matrix or i-th categorical field, $v_i$ and $d$ is the vocabulary size and embedding size. Numerical features are used both as dense features by Non-discretization approach and embedding by Discretization approach. For the j-th numerical feature field, the feature embedding can be obtained by a two-stage operation: discretization and embedding look-up.
\begin{equation}
    \bm{e}^{n}_j=\bm{E}^{n}_j\cdot onehot(d_j(x^{n}_i))
\end{equation}
where $\bm{E}^{n}_j \in \mathbb{R}^{w_i \times d}$ is the learnable embedding matrix or j-th numerical field, $w_i$ is the number of the buckets after discretization. Then all the embedding of categorical features and numerical features, as well as dense values of numerical features are concatenated to form the input of MLP Module afterwards, i.e,
\begin{equation}
\begin{split}
    \bm{z}=concatenate([\bm{e}^{c}_1,\bm{e}^{c}_2,...,\bm{e}^{c}_M,\\
            \bm{e}^{n}_1,\bm{e}^{n}_2,...,\bm{e}^{n}_N,\\
            [x^{n}_{1},x^{n}_{2},...,x^{n}_{N}]])
\end{split}
\end{equation}
Subsequently, the estimated CTR $\hat{y}$ can be obtained by the following discriminative model,
\begin{equation}
\hat{y}=\sigma(f_{\bm{\theta}}(\bm{z}))
\label{predict}
\end{equation}
where $f_{\bm{\theta}}(\cdot)$ refers to the function of the MLP Module which is parameterized by $\bm{\theta}$, and $\sigma(\cdot)$ is the sigmoid activation function. 
The model parameters $\bm{E}$ and $\bm{\theta}$ are learned by maximizing the objective function $\mathcal{L}(\bm{E},\bm{\theta})$ with gradient-based optimization methods.
In the traditional point estimate CTR prediction model, the objective function is equal to the negative log-likelihood $l(\bm{E},\bm{\theta})$:
\begin{equation}
\begin{aligned}
\mathcal{L}(\bm{E},\bm{\theta}) &= l(\bm{E},\bm{\theta}) \\
&\equiv -y {\rm log}\hat{y}-(1-y) {\rm log}(1-\hat{y})
\end{aligned}
\end{equation}

\begin{figure*}[t]
\centering
\includegraphics[width=1.0\linewidth]{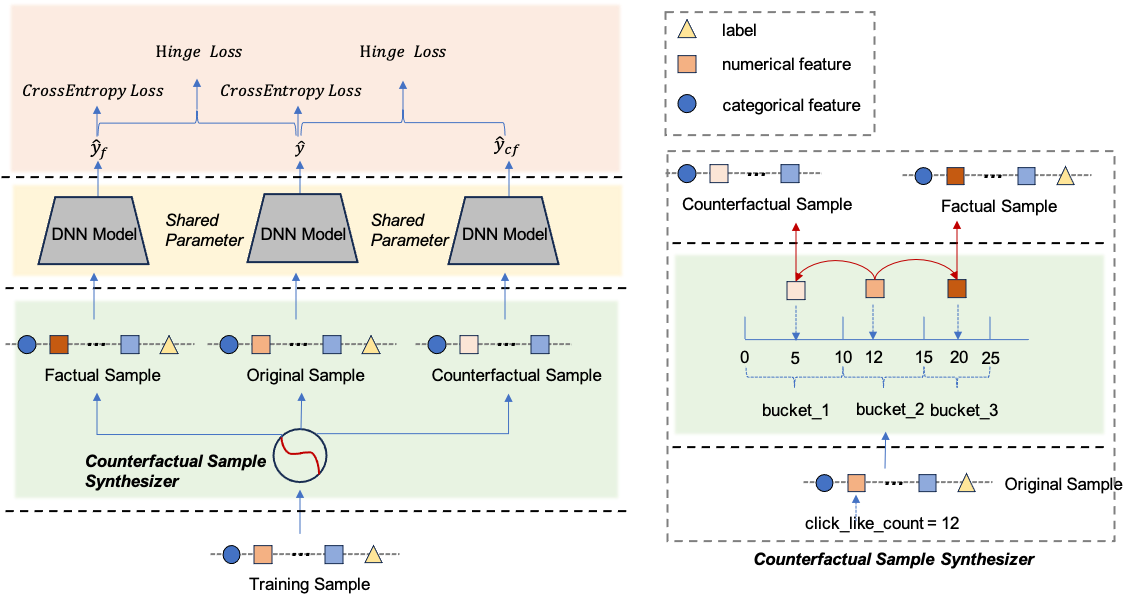}
\caption{An illustration of our proposed CCSS: the left is the end-to-end learning framework with counterfactual sample synthesizing, and the right is the detail illustration of the Counterfactual Sample Synthesizer.}
\label{illu_main_model}
%\vspace{-1.0em}
\end{figure*}

\subsection{Counterfactual Sample Synthesizing}
% 给定一个原始样本，如何得到factual sample和counterfactual sample\
To take advantage of the contrastive loss to the greatest extent, we synthesize generated samples as realistic as possible, i.e., the difference between generated samples and the original samples are as small as possible.
Therefore, we disturb only one numerical feature for each original sample.
To generate samples as informative as possible, we introduce feature importance as the specific metric to quantize the probability to be disturbed for each numerical feature.
In this chapter, we describe the implementation details to synthesize realistic and informative generated samples and how to train with them, which is illustrated in Figure \ref{illu_main_model}.

\subsubsection{Interpreting Feature Importance}
In machine learning, feature importance is a specific metric to measure the marginal contribution of each feature to model's decisions. When the model's decisions are affected much stronger by a specific feature, the stronger interpretability between our model's outputs and this specific feature is expected. To try to conform the expectation, we introduce feature importance during synthesizing counterfactual samples. We interpret feature importance of each numerical feature using Shapley Value, which is one of the most widely adopted measures of feature importance as it has a solid theoretical foundation \cite{song2016shapley, fryer2021shapley}.

\subsubsection{Feature Value Disturbing}
% random sample
We synthesize one counterfactual sample and one factual sample by disturbing one of the numerical features of each original samples. The synthesizing process can be divided into the following steps:

1) Obtain the feature importance $q_i$ for each numerical feature where $i$ is the index of the numerical feature.

2) Calculate the probability $p_i = \frac{q_i}{\sum_{j=1}^{N}q_j}$ to be disturbed for each numerical feature, where $i$ is the index of the numerical feature.

3) Given a training sample, choose the index of the numerical feature to be disturbed by drawing floats from the uniform distribution over $[0,1)$.

4) For an original training sample with positive label, we synthesize one counterfactual sample $\mathcal{C}$ by disturbing the feature value to the center of the left neighbor bucket while keeping other features unchanged, and the counterfactual sample has no known label. At the same time, we synthesize one factual sample $\mathcal{F}$ by disturbing the feature value to the center of the right neighbor bucket while keeping other features unchanged, and the factual sample has positive label. On the contrary, for an original training sample $\mathcal{O}$ with negative label, one counterfactual sample without known label is synthesized by disturbing the feature value to the center of the right neighbor bucket, and one factual sample with negative sample is synthesized by disturbing the feature value to the center of the left neighbor bucket. Here, we take the monotonic increasing relationship between model output and numerical inputs as an illustration. Table 
\ref{tab:ccss} demonstrates the monotonic decreasing scenario in detail.

\begin{table}[t]
  \centering
  \caption{An detailed illustration of counterfactual samples synthesizing.}
  \label{tab:ccss}
  \vspace{0.5em}
  \begin{tabular}{c|c|cc}
    % \toprule
    \hline
    \multirow{2}{*}{Monotonicity}& \multirow{2}{*}{Label}& \multicolumn{2}{c}{Disturb Destination}\\
    % \midrule
     & & Counterfactual & Factual\\
    % \midrule
    \hline
    % \midrule
    \multirow{2}{*}{Increasing}
        & Positive & Right & Left\\
        & Negative & Left & Right\\
    % \midrule
    \hline
    % \midrule
    \multirow{2}{*}{Decreasing}
        & Positive & Left & Right\\
        & Negative & Right & Left\\
    
    % \bottomrule
    \hline
    \end{tabular}
\vspace{-1.0em}
\end{table}

\textbf{The Boundary Condition.} For an original training sample with positive label, when it's feature value to be disturbed locates in the far-right bucket, we only synthesize one counterfactual sample by disturbing the feature value to the left neighbor bucket. Similarly, for an original training sample with negative label, when it's feature value to be disturbed locates in the far-left bucket, we only synthesize one counterfactual sample by disturbing the feature value to the right neighbor bucket.

% disturbed to move to the right neighbor bucket

% disturbed to move to the left neighbor bucket

% boundary conditions

\subsection{Training with Counterfactual Samples}
In this chapter, we describe the training implementation details with the synthesized samples and the original samples, which is illustrated in Figure \ref{illu_main_model}.
% Loss的重新设计
\subsubsection{Learning to Contrast}
Here, we also take the monotonic increasing relationship between model output and numerical inputs as an illustration.
For a training instance with positive label, the monotonicity means that we expect that its synthesized factual sample $\mathcal{F}$ is scored higher by the model than this original sample $\mathcal{O}$, and this original sample is scored higher than its synthesized counterfactual sample $\mathcal{C}$.
To address the monotonicity expectation in the CTR prediction model, we add pairwise losses to contrast the $(\mathcal{F},\mathcal{O})$ pair and the $(\mathcal{O},\mathcal{C})$ pair:
\begin{equation}
\begin{aligned}
\mathcal{L}(\bm{E},\bm{\theta}) &= l(\bm{E},\bm{\theta}) \\
&+ \alpha (l^{\mathcal{P}}(\bm{E},\bm{\theta},\mathcal{F},\mathcal{O}) \\
&+l^{\mathcal{P}}(\bm{E},\bm{\theta},\mathcal{O},\mathcal{C}))
\end{aligned}
\end{equation}
where $l^{\mathcal{P}}$ denotes the pairwise loss, and hinge loss is adopted in this paper. we introduce a hyper\-parameter $\alpha$ to control the trade-off between pairwise loss and point-wise loss during training.
Similarly, for a training instance with negative label, we expect that its synthesized counterfactual sample $\mathcal{C}$ is scored higher than this original sample $\mathcal{O}$, and this original sample $\mathcal{O}$ is scored higher than its synthesized factual sample $\mathcal{F}$. Therefore, loss function for a training instance $\mathcal{O}$ with negative label is formulized as:
\begin{equation}
\begin{aligned}
\mathcal{L}(\bm{E},\bm{\theta}) &= l(\bm{E},\bm{\theta}) \\
&+ \alpha (l^{\mathcal{P}}(\bm{E},\bm{\theta},\mathcal{O},\mathcal{F}) \\
&+l^{\mathcal{P}}(\bm{E},\bm{\theta},\mathcal{C},\mathcal{O}))
\end{aligned}
\end{equation}

\subsubsection{Data Augmentation with Factual Samples}
According to the priori monotonicity, the synthesized factual samples have known labels. For example, with the priori increasing monotonicity, a synthesized factual sample for a training instance with positive label is known to be labeled positive, while to be labeled negative for a training instance with negative label. In this paper, we exploit these synthesized factual sample with known labels to augment training data. Hence, our loss function can be rewritten as:
\begin{equation}
\begin{aligned}
\mathcal{L}(\bm{E},\bm{\theta}) &= l(\bm{E},\bm{\theta},\mathcal{O}) \\
&+ l(\bm{E},\bm{\theta},\mathcal{F}) \\
&+ \alpha (l^{\mathcal{P}}(\bm{E},\bm{\theta},\mathcal{O},\mathcal{F}) \\
&+l^{\mathcal{P}}(\bm{E},\bm{\theta},\mathcal{C},\mathcal{O}))
\end{aligned}
\end{equation}

\section{Experiments}
In this section, we conduct experiments with the aim of answering the following three research questions:
\begin{itemize}
    \item[\textbf{RQ1}] How does CCSS preform for improving the interpretability and effectiveness of deep models in recommender?
    % vs. 非bayesian的退化版本，given normal gaussian priors
    \item[\textbf{RQ2}] How does CCSS perform when plugged into various network backbones?
    \item[\textbf{RQ3}] What are the effects of the random strategy for selecting feature to be disturbed, constrastive loss, data augmentation loss and hyper-parameter $\alpha$ in CCSS?
\end{itemize}

\subsection{Dataset}
% We evaluate the performance of our proposed approaches on three publicly available datasets and a real industrial dataset:
We evaluate the performance of our proposed approaches on a public benchmark dataset and our real industrial dataset. 

\textbf{KuaiRand-Pure\footnote{\url{https://github.com/chongminggao/KuaiRand}}:} 
This data set is collected from the recommendation logs of the video-sharing mobile app, Kuaishou. It contains numerous numerical features indicating the statistical information of user behaviors and video behaviors. We adopt $is\_click$ as the binary label and adopt its full feature set for training and testing. Samples before 05-01 are used for training, and samples during 05-01 and 05-08 are used for testing.

\textbf{Real Industrial Dataset:} 
This data set is sampled and collected from the realtime data stream of our online video-sharing platform. It consists of 1.29 billion video play records generated by 43.98 million users within 8 days.
Features in this dataset contain user ID, video ID and 14 dimensional numerical features which contain rich user activity level, video popularity and user-item interaction activity level information. The feature description in our real industrial dataset is shown in Table \ref{fea-desc}. Each instance in this dataset contains a binary label $is\_collect$ indicates whether the user collect this video to her favorites list in this instance. Our experiments are conducted on a binary classification model to predict $is\_collect$.
Samples from the beginning 7 days are used for training, and samples from the following 8-th day are used for testing.
The statistics of the real industrial dataset can be found in Table \ref{stat-data}.

For both KuaiRand and our real industrial dataset, there is a strong semantic correlation between the input numerical features and the estimated target, and we expect that "the larger the input value, the larger the output result should be".

\begin{table}[h]
  \centering
  \caption{Feature description for our industrial dataset.}
  \label{fea-desc}
  \vspace{0.1em}
  \renewcommand\arraystretch{2.0}
  \begin{tabular}{|m{2.2cm}|m{5.0cm}|}
    % \toprule
    \hline
    \textbf{Feature Types}&  \textbf{Descriptions}\\
    % \midrule
    \hline
    % \midrule
    \textbf{User ID}
        & the unique identification of the user.\\
    % \midrule
    \hline
    % \midrule
    \textbf{Video ID}
        & the unique identification of the video.\\
    % \midrule
    \hline
    % 全局的emprical_xtr
    % \textbf{ctr\_global, ltr\_global, wtr\_global, ftr\_global, vtr\_global, lvtr\_global, cmtr\_global } 
    \textbf{video\_ctr, video\_ltr, video\_wtr, video\_ftr, video\_vtr, video\_lvtr, video\_cmtr }
        & global empirical click rate, like rate, follow rate, forward rate, video viewing duration, longview rate, comment rate of this video during 30days.\\
    % \midrule
    \hline
    % 当前用户 x 当前类目 empirical xtr
    \textbf{user\_ctr, user\_ltr, user\_wtr, user\_ftr, user\_vtr, user\_lvtr, user\_cmtr }
        & global empirical click rate, like rate, follow rate, forward rate, video viewing duration, longview rate, comment rate of this user on this video's category during 30days.\\
    % \bottomrule
    \hline
    \end{tabular}
\vspace{-1.0em}
\end{table}

\begin{table}[h]
  \centering
  \caption{Statistics of the training dataset.}
  \label{stat-data}
  \vspace{0.1em}
  \renewcommand\arraystretch{1.5}
  \begin{tabular}{c|c}
    \hline
    % \midrule
    {\#user}
        & 43.98 Mil.\\
    % \midrule
    \hline
    % \midrule
    {\#video}
        & 23.80 Mil.\\
    % \midrule
    \hline
    % \midrule
    {\#training sample}
        & 1.29 Bil.\\
    % \midrule
    \hline
    % \midrule
    {\#test sample}
        & 0.26 Bil.\\
    % \midrule
    % \hline
    % \bottomrule
    \hline
    \end{tabular}
%\vspace{-1.0em}
\end{table}

\begin{table*}[tp]
  \centering
  \caption{Model comparison on KuaiRand and our industrial dataset. We record the mean results over 5 runs. Std $\approx$ 0.1\%, extremely statistically significant under unpaired t-test. * indicates the improvement is statistically significant at the significance level of 0.05 over the best baseline on AUC or GAUC. `$\mathcal{K}$th\_fea' is short for `Mono\_rate of the numerical feature who has the $\mathcal{K}$th highest feature importance score'.}
  %\vspace{-1.0em}
  \label{main-tab}
  \begin{tabular}{cc||cc||cc||c|c|c|c|c|c|c}
    \toprule
    \multirow{11}{*}{\rotatebox{90}{Our Industrial Dataset}} & Models & AUC & RelaImpr & GAUC & RelaImpr & 1st\_fea & 2nd\_fea & 3rd\_fea & 4th\_fea & 5th\_fea & 6th\_fea & 7th\_fea\\
    \specialrule{0em}{1pt}{1pt}
    \cline{2-13}
    \specialrule{0em}{1pt}{1pt}
    & DNN & 0.7200 & - & 0.6560 & - & 51.3\% & 59.8\% & 53.7\% & 53.9\% & 54.5\% & 55.0\% & 59.6\%
 \\
    & DNN +CCSS & 0.7860$^{*}$ & +30.0\% & 0.7320$^{*}$  & +48.7\% & 83.6\% & 91.7\% & 79.2\% & 65.7\% & 82.7\% & 83.9\% & 87.3\%
 \\
    \specialrule{0em}{1pt}{1pt}
    \cline{2-13}
    \specialrule{0em}{1pt}{1pt}
    & Wide \& Deep & 0.7311 & - & 0.6672 & - & 54.5\% & 63.9\% & 55.8\% & 51.7\% & 50.1\% & 45.7\% & 41.0\%
 \\
    & Wide \& Deep +CCSS & 0.7763$^{*}$ & +19.6\% & 0.7110$^{*}$ & +26.2\% & 87.5\% & 89.1\% & 82.0\% & 76.4\% & 74.9\% & 87.3\% & 78.9\%
 \\
    \specialrule{0em}{1pt}{1pt}
    \cline{2-13}
    \specialrule{0em}{1pt}{1pt}
    & PNN & 0.7253 & - & 0.6662 & - & 52.6\% & 54.1\% & 56.9\% & 58.1\% & 48.2\% & 52.9\% & 55.7\%
 \\
    & PNN +CCSS & 0.7663$^{*}$ & +18.2\% & 0.6934$^{*}$ & +16.4\% & 88.0\% & 83.9\% & 80.3\% & 78.9\% & 70.2\% & 74.7\% & 87.9\%
 \\
    \specialrule{0em}{1pt}{1pt}
    \cline{2-13}
    \specialrule{0em}{1pt}{1pt}
    & DCN & 0.7356 & - & 0.6702 & - & 53.7\% & 56.5\% & 50.4\% & 46.7\% & 47.8\% & 47.5\% & 49.5\%
 \\
    & DCN +CCSS & 0.7865$^{*}$ & +21.6\% & 0.7242$^{*}$ & +31.7\% & 87.9\% & 90.2\% & 88.9\% & 88.0\% & 78.2\% & 87.5\% & 85.0\%
 \\ 
    \specialrule{0em}{1pt}{1pt}
    \cline{2-13}
    \specialrule{0em}{1pt}{1pt}
    & DeepFM & 0.7168 & - & 0.6545 & -	& 42.7\% & 45.6\% & 53.2\% & 53.6\% & 48.7\% & 53.8\% & 51.6\%
 \\
    & DeepFM +CCSS & 0.7298$^{*}$ & +6.0\% & 0.6615$^{*}$ & +4.5\%	& 81.1\% & 83.2\% & 80.8\% & 73.5\% & 74.0\% & 87.9\% & 84.5\%
 \\
    \midrule
    \multirow{11}{*}{\rotatebox{90}{KuaiRand-Pure}} & Models & AUC & RelaImpr & GAUC & RelaImpr & 1st\_fea & 2nd\_fea & 3rd\_fea & 4th\_fea & 5th\_fea & 6th\_fea & 7th\_fea\\
    \specialrule{0em}{1pt}{1pt}
    \cline{2-13}
    \specialrule{0em}{1pt}{1pt}
    & DNN & 0.6847 & - & 0.6419 & - & 60.3\% & 58.2\% & 49.7\% & 50.9\% & 54.5\% & 50.2\% & 40.5\%
 \\
    & DNN +CCSS & 0.7324$^{*}$ & +25.8\% & 0.6768$^{*}$ & +24.5\% & 66.5\% & 62.1\% & 65.1\% & 71.3\% & 80.5\% & 66.7\% & 75.8\%
 \\
    \specialrule{0em}{1pt}{1pt}
    \cline{2-13}
    \specialrule{0em}{1pt}{1pt}
    & Wide \& Deep & 0.7113 & - & 0.6626 & - & 61.5\% & 59.7\% & 51.3\% & 51.5\% & 57.8\% & 51.1\% & 45.7\%
 \\
    & Wide \& Deep +CCSS & 0.7527$^{*}$ & +19.6\% & 0.6959$^{*}$ & +20.5\% & 68.0\% & 65.4\% & 64.6\% & 72.7\% & 88.3\% & 68.8\% & 75.2\%
 \\
    \specialrule{0em}{1pt}{1pt}
    \cline{2-13}
    \specialrule{0em}{1pt}{1pt}
    & PNN & 0.7222 & - & 0.6702 & - & 63.6\% & 60.9\% & 52.2\% & 52.0\% & 55.5\% & 50.6\% & 46.4\%
 \\
    & PNN +CCSS & 0.7491$^{*}$ & +12.1\% & 0.7039$^{*}$ & +19.8\% & 68.6\% & 74.7\% & 63.3\% & 79.2\% & 87.3\% & 70.8\% & 77.5\%
 \\
    \specialrule{0em}{1pt}{1pt}
    \cline{2-13}
    \specialrule{0em}{1pt}{1pt}
    & DCN & 0.7086 & - & 0.6605 & - & 59.3\% & 59.2\% & 49.7\% & 55.9\% & 53.5\% & 52.2\% & 47.5\%
 \\
    & DCN +CCSS & 0.7737$^{*}$ & +31.2\% & 0.6873$^{*}$ & +16.7\% & 74.5\% & 75.8\% & 70.1\% & 78.0\% & 78.3\% & 73.2\% & 80.3\%
 \\ 
    \specialrule{0em}{1pt}{1pt}
    \cline{2-13}
    \specialrule{0em}{1pt}{1pt}
    & DeepFM & 0.6912 & - & 0.6476 & -	& 60.8\% & 58.4\% & 51.5\% & 51.1\% & 52.3\% & 51.8\% & 46.9\%
 \\
    & DeepFM +CCSS & 0.7357$^{*}$ & +23.3\% & 0.6846$^{*}$ & +25.1\%	& 65.8\% & 62.7\% & 67.2\% & 78.9\% & 78.7\% & 69.1\% & 78.6\%
 \\
    \bottomrule
  \end{tabular}
\vspace{-1em}
\end{table*}

\subsection{Baselines}
In order to prove the plug-and-play nature of the method, we did sufficient tests on different backbones:

(1) DNN is the lightest model structure which only contains a MLP and some nonlinear activation functions.

(2) Wide \& Deep \cite{cheng2016wide} develop wide linear models and deep neural networks together to enhance their respective abilities.

(3) DeepFM \cite{guo2017deepfm} is a deep recommendation method that learns both low and high level interactions between fields.  

(4) DCN \cite{wang2017deep} is based on DNN, explicitly applies feature crossing at each layer, eliminating the need for human feature engineering. 

(5) PNN \cite{qu2016product} employs a feature extractor to investigate feature interactions among inter-field categories.

\subsection{Experimental Settings}
In this chapter, we describe the implementation details and the evaluation metrics in our experiments.

\subsubsection{Implementation Details}
We utilize the same model settings for all approaches on each dataset to provide a fair comparison. For all the three datasets, we fix embedding size as 32 and DNN as 4 FC layers with [512, 255, 127, 127] hidden units. Furthermore, for DCN, we set the number of cross layer to 3. We optimize all approaches using mini-batch Adam, where the learning rate is 0.05 and decay\_rate is 0.9. Furthermore, the batch size of all models is set to 1024.
We adopt hyper-parameter $\alpha=1.0$ for all the offline and online evaluations. 

\subsubsection{Evaluation Metrics} We adopt \textbf{AUC} and \textbf{Mono\_rate} to evaluate the effectiveness and interpretability of our model.

\textbf{AUC:} \cite{auc} is a common metric for recommendation \cite{wang2017deep,zhou2018deep}. It measures the goodness of order by ranking all the items with prediction. Thus we adopt AUC as the main metric. 

\textbf{GAUC:} gauc is introduced in \cite{zhou2018deep} which measures the goodness of intra-user order by averaging AUC over users and is shown to be more relevant to online performance in recommendation system. We adapt this metric in our experiments, gauc is defined as:
    \begin{equation}
    GAUC=\frac{ {\textstyle \sum_{i=1}^{n}}  \#impression_{i} * AUC_{i}}{{\textstyle \sum_{i=1}^{n}}  \#impression_{i}}
    \end{equation}
where n is the number of users, $\rm{\#impression}_i$ and $\rm{AUC}_i$ are the number of impressions and AUC corresponding to the i-th user.

In addition, we follow \cite{wang2017deep,zhou2018deep} to introduce RelaImpr metric to measure relative improvement over models. For a random guesser, the value of AUC or GAUC is 0.5. Hence, RelaImpr is defined as:
    \begin{equation}
    RelaImpr(AUC)=(\frac{AUC(measured\ model)-0.5}{AUC(base\ model)-0.5}-1) \times 100\%
    \end{equation}

    \begin{equation}
    RelaImpr(GAUC)=(\frac{GAUC(measured\ model)-0.5}{GAUC(base\ model)-0.5}-1) \times 100\%
    \end{equation}

    % vs. 非bayesian的退化版本，given normal gaussian priors
\textbf{Mono\_rate:} We define Mono\_rate to evaluate the monotonicity (i.e., interpretability) between model output and numerical features:
    \begin{equation}
    \begin{aligned}
    Mono\_Rate = \frac{\#Monotone\_pairs(D)}{\#Comparable\_pairs(D)}
    \end{aligned}
    \end{equation}
where $Comparable\_pairs(D)$ donates the entire set of the expected monotonic pairs, including all the $(\mathcal{F},\mathcal{O})$ pairs and $(\mathcal{C},\mathcal{O})$ pairs. $Monotone\_pairs(D)$ donates the set of valid predicted monotonic pairs according to model outputs. For an original sample with positive label, the valid predicted monotonic pairs include $(\mathcal{F},\mathcal{O})$ pairs whose $\hat{y}_f>\hat{y}$ and $(\mathcal{C},\mathcal{O})$ pairs whose $\hat{y}_cf<\hat{y}$. Instead, for an original sample with negative sample, the valid predicted monotonic pairs include $(\mathcal{F},\mathcal{O})$ pairs whose $\hat{y}_f<\hat{y}$ and $(\mathcal{C},\mathcal{O})$ pairs whose $\hat{y}_{cf}>\hat{y}$.

\subsection{Comparison with Baselines (RQ1, RQ2)}

\subsubsection{Offline Evaluation}
To demonstrate the effectiveness and adaptability of the our proposed CCSS, we plug CCSS in many representative networks in binary classification task, such as DNN, Wide\&Deep, DCN, DeepFM and PNN. According to Table \ref{main-tab}, CCSS could improve all of the baselines on two large scale dataset. CCSS brings at least 6.0\% AUC improvement and 4.5\% GAUC improvement. 
To evaluate the effectiveness of CCSS to improve the monotonicity between model output and the numerical features during infer, we calculate Mono\_rates with testing dataset. For simplicity, we calculate and compare the Mono\_rates of the top 7 important numerical features on KuaiRand and our industrial dataset. According to Table \ref{main-tab}, Mono\_rate improvements are clearly remarkable, which means the interpretability of the neural network can be enhanced by our proposed CCSS.

\subsubsection{Online Evaluation}
To exam the effectiveness of CCSS in a real industrial recommendation scenario, we conduct online A/B testing based on a $is\_collect$ predicting model. Table \ref{ab-test} shows the experimental results. 
CCSS contributes 3.93\% $collect\_rate$ gain when plugged into DCN which has already been highly optimized for our online system with abundant of features.

\begin{table}[H]
  \centering
  \caption{Online A/B Testing.}
  \label{ab-test}
  \renewcommand\arraystretch{1.5}
    \begin{tabular}{l|c}
    \hline
    Online A/B Test & collect\_rate gain\\
    % \midrule
    \hline
    DCN & -\\
    DCN + CCSS & 3.93\% \\
    % \bottomrule
    \hline
    \end{tabular}
  %\vspace{-2em}
\end{table}

%\textcolor{red}{TODO: add future work in conclusion, add another dataset, add some discussion}

\subsection{Ablation Study(RQ3)}
In this section, we demonstrate the advantages of our proposed random strategy for selecting feature to be disturbed, constrastive loss, data augmentation loss in CCSS. 
We present an ablation study on our real industrial dataset by evaluation several models based on DNN which is the lightest structures:
(1) CCSS(Only Data Augmentation): only adopt the auxiliary point-wise loss of factual samples.
(2) CCSS(Only Factual Contrastive Loss): only adopt the auxiliary hinge loss of Factual-Original sample pairs.
(3) CCSS(Only Counterfactual Contrastive Loss): only adopt the auxiliary hinge loss of Counterfactual-Original sample pairs.
(3) CCSS(Equal Probability Random Strategy): select the numerical feature to be disturbed with equal probability.
The mean AUC and GAUC results over 5 runs are reported in Table \ref{ablation}, and the results confirm the advantages of our proposed random strategy for selecting feature to be disturbed, constrastive loss, data augmentation loss in CCSS.

\begin{figure}[t]
\centering
    \begin{minipage}[t]{0.8\linewidth}
      \centering
      %\centerline{\includegraphics[width=1\linewidth]{base-effect-2.pdf}}
      \centerline{\includegraphics[width=0.75\linewidth]{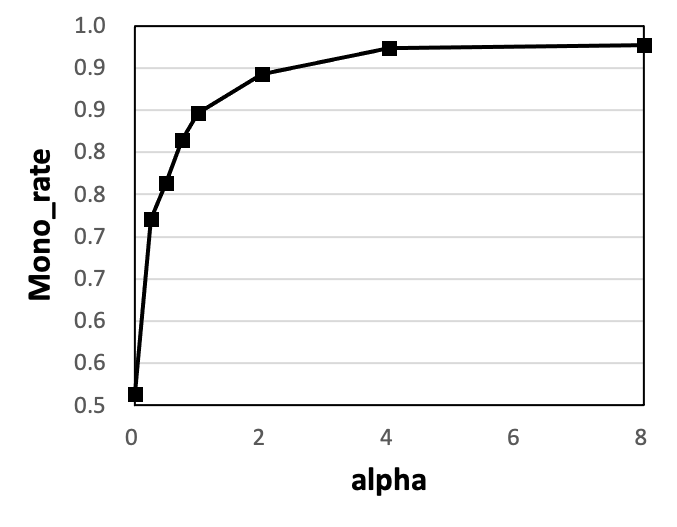}}
      \vspace{-0.5em}
      \centerline{\footnotesize{(a)}}
    %   \vspace{-0.5em}
      \centering
    \end{minipage}%
    \hspace{0.1em}
    \begin{minipage}[t]{0.8\linewidth}
      \centering
      %\centerline{\includegraphics[width=1\linewidth]{base-effect.pdf}}
      \centerline{\includegraphics[width=0.75\linewidth]{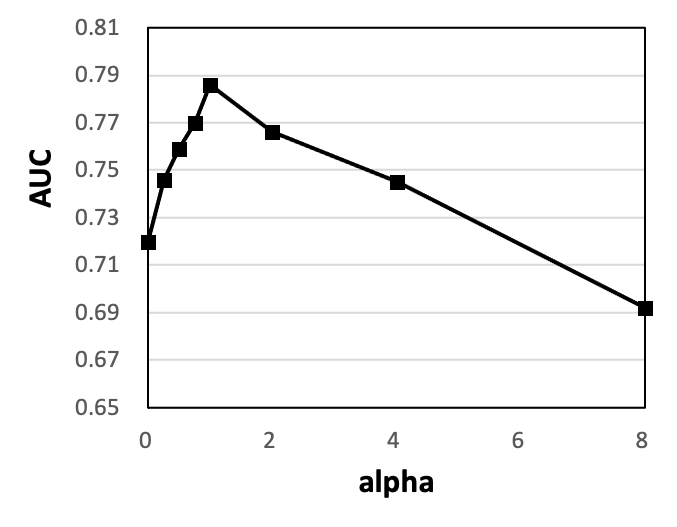}}
      \vspace{-0.5em}
      \centerline{\footnotesize{(b)}}
    %   \vspace{-0.5em}
      \centering
    \end{minipage}%
\vspace{-1.0em}
\caption{Effect of $\alpha$ on (a) Mono\_rate and (b) AUC. DNN is used as the network.}
\label{hyper-param}
%\vspace{1em}
\end{figure}

\textbf{Effect of Hyper\-Parameter $\alpha$.}
We evaluate the effect of hyper-parameter $\alpha$ which controls the trade-off between pairwise loss and point-wise loss.
According to Figure \ref{hyper-param}, contrastive losses aimed to improve the interpretability of the model, can help to promote the effectiveness of the model with an appropriate proportion.

\begin{table}[H]
  \centering
  \caption{Ablation study. The averaged AUC and GAUC results over 5 runs are reported. }
  \label{ablation}
  %\vspace{0.5em}
  \renewcommand\arraystretch{1.5}
  \begin{tabular}{c|c|c}
    % \toprule
    \hline
    % \multirow{2}{*}{Model}& \multicolumn{2}{c}{AUC} & \multicolumn{2}{c}{AUC} & \multicolumn{2}{c}{AUC}\\
    % \midrule
     Models & AUC & GAUC \\
    % \midrule
    \hline
    % \midrule
    \multirow{1}{*}{DNN}
        & 0.786 & 0.732 \\
    % \midrule
    \hline
    % \midrule
    \multirow{1}{*}{DNN(Only Factual Pairwise loss)}
        & 0.772 & 0.715 \\
    % \midrule
    \hline
    % \midrule
    \multirow{1}{*}{DNN(Only Counterfactual Pairwise loss)}
        & 0.768 & 0.720 \\
    % \midrule
    \hline
    % \midrule
    \multirow{1}{*}{DNN(Equal Probability Random Disturb)}
        & 0.766 & 0.711 \\
    % \midrule
    \hline
    % \midrule
    \multirow{1}{*}{DNN(Only Factual Pointwise loss)}
        & 0.734 & 0.669 \\
    % \bottomrule
    \hline
    \end{tabular}
%\vspace{-1.0em}
\end{table}

\section{Conclusion}
In this paper, we propose a general model-agnostic Contrastive learning framework with Counterfactual Samples Synthesizing (CCSS) for modeling the monotonicity between the neural network output and numerical features. CCSS introduces and models the monotonicity by synthesizing counterfactual sample and factual samples and learning to contrast among the counterfactual sample, factual samples and original samples. Besides, to generate samples as informative as possible, we introduce feature importance as the specific metric to quantize the probability to be disturbed for each numerical feature. Our proposed CCSS can be adapted to various representative networks without much effort and enjoys an end-to-end learning manner. Abundant offline and online experiments show that recommendation neural networks with CCSS can achieve better interpretability and performances.

\bibliographystyle{ACM-Reference-Format}
\balance
\bibliography{main}

\end{document}